\documentclass[conference]{IEEEtran}
\IEEEoverridecommandlockouts
\usepackage{cite}
\usepackage{amsmath,amssymb,amsfonts}
\usepackage{algorithmic}
\usepackage{graphicx}
\usepackage{textcomp}
\usepackage{xcolor}
\def\BibTeX{{\rm B\kern-.05em{\sc i\kern-.025em b}\kern-.08em
    T\kern-.1667em\lower.7ex\hbox{E}\kern-.125emX}}
\usepackage{tikz}

\usepackage{microtype}
\usepackage{caption}
\usepackage{subcaption}
\usepackage{pgfplots}
\pgfplotsset{compat=1.18}
\usepgfplotslibrary{polar}
\usepackage{ifthen}
\usepackage{comment}
\usepackage{array}
\usepackage[capitalise]{cleveref}
\definecolor{matlab1}{rgb}{0,0.4470,0.7410}
\definecolor{matlab2}{rgb}{0.8500,0.3250,0.0980}
\definecolor{matlab3}{rgb}{0.9290,0.6940,0.1250}
\definecolor{matlab4}{rgb}{0.4940,0.1840,0.5560}
\definecolor{matlab5}{rgb}{0.4660,0.6740,0.1880}
\definecolor{matlab6}{rgb}{0.3010,0.7450,0.9330}
\definecolor{matlab7}{rgb}{0.6350,0.0780,0.1840}
\renewcommand\vec{\mathbf}

\usepackage[font = footnotesize, textfont=up]{caption}
\usepackage{tikz}

\newcommand\copyrighttext{%
  \footnotesize \textcopyright \the\year{} IEEE. Personal use of this material is permitted. Permission from IEEE must be obtained for all other uses, including reprinting/republishing this material for advertising or promotional purposes, collecting new collected works for resale or redistribution to servers or lists, or reuse of any copyrighted component of this work in other works.}

\newcommand\copyrightnotice{%
\begin{tikzpicture}[remember picture,overlay]
\node[anchor=south,yshift=10pt] at (current page.south) {\fbox{\parbox{\dimexpr0.75\textwidth-\fboxsep-\fboxrule\relax}{\copyrighttext}}};
\end{tikzpicture}%
}

\begin{document}
\bstctlcite{IEEEexample:BSTcontrol}

\title{DoA Estimation with Sparse Arrays: Effects of Antenna Element Patterns and Nonidealities}

\author{\IEEEauthorblockN{
Niko Lindvall\IEEEauthorrefmark{1},
Mikko Heino\IEEEauthorrefmark{1},
Robin Rajamäki\IEEEauthorrefmark{2},
Mikko Valkama\IEEEauthorrefmark{1},
and Visa Koivunen\IEEEauthorrefmark{2}
}                                     %
\IEEEauthorblockA{%
\IEEEauthorrefmark{1}Tampere Wireless Research Center, Unit of Electrical Engineering, %
Tampere University, Finland}
\IEEEauthorrefmark{2}School of Electrical Engineering, %
Aalto University, Finland %
\vspace{-3mm}
}
\maketitle
\copyrightnotice
\begin{abstract}
This paper studies the effects of directional antenna element complex gain patterns and nonidealities in direction of arrival (DoA) estimation. We compare sparse arrays and classical uniform linear arrays, harnessing EM simulation tools to accurately model the electromagnetic behavior of both patch and Vivaldi antenna element including mutual coupling effects. We show that with sparse array configurations, the performance impacts are significant in terms of DoA estimation accuracy and operable SNR ranges. Specifically, in the scenarios considered, both the usage of directional antenna elements and a sparse array result in over 90\% reduction in average direction finding error, compared to a uniform omnidirectional array with the same number of elements (in this case eight), when estimating the directions of two sources using the MUSIC algorithm. For a fixed angular RMSE, the improvements in array sensitivity are shown to yield a 4 to 15-fold increase in one-way coverage distance (assuming free-space path loss). Among the studied options, the best performance was obtained using sparse arrays with either patch or Vivaldi elements for field of views of 100$^\circ$ or 120$^\circ$, respectively.

\end{abstract}

\begin{IEEEkeywords}
antenna pattern, %
DoA estimation, nonidealities, patch element, sparse arrays, Vivaldi element, EM-simulation
\end{IEEEkeywords}

\vspace{-1mm}
\section{Introduction}

\textls[-1]{\textit{Direction-of-arrival} (DoA) estimation is a crucial task of many positioning, radar, wireless communications, and spectrum monitoring sensor array systems. %
The key factors affecting the resolution of DoA estimation are the geometry and electrical aperture size of the sensor array. The most commonly used array configuration is %
the \textit{uniform linear array} (ULA) \cite{vantrees2002optimum}. However, increasing the aperture size of ULAs leads to higher costs and increased use of resources because a large number of sensors and related circuitry are required.
\textit{Sparse arrays} enable significantly reducing the number of sensors while still maintaining the same aperture size and resolution, or increasing the aperture size and resolution while maintaining the same number of elements \cite{amin2024sparse}. Sparse arrays also come with the additional benefit that they can resolve more sources than the number of sensors \cite{coarray_music}.%
}

\textls[-8]{Typically, sparse array DoA estimation has been studied with ideal isotropic antennas \cite{amin2024sparse}. Only few articles discuss the effects of real-world antenna elements with practical element responses and other potential nonidealities due to for example manufacturing tolerances. 
}
\textls[-8]{In \cite{array_element_pattern_in_doa}, the effect of the antenna element gain pattern on ULAs employing high-resolution DoA estimation algorithms, including ESPRIT and MUSIC, was studied. 
Array processing in presence of nonidealities using arbitrary array configurations was discussed in \cite{manifold_separation}. 
The impact of nonidealities or model mismatch on the performance of various linear arrays employing MUSIC has also been investigated numerically \cite{coupling,music_location_errors,friedlander2018antenna} 
and experimentally \cite{wang2018experimental,laakso2021phase,zheng2024enhancing}. 
Further experimental results have been obtained using 
compact reconfigurable linear leaky wave antennas in \cite{Experimental_DOA_compact_reconfigurable}. 
A four-element coprime array was also compared to a 
ULA 
with monopole and patch elements in \cite{Sparse_DOA_experimental}.}
\begin{figure}
    \centering
     \centering
     \begin{subfigure}[b]{0.5\textwidth}
         \centering
        \includegraphics{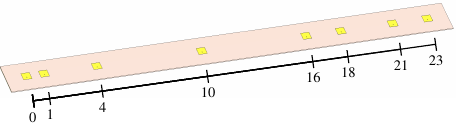}
        \vspace{-0.5cm}
         \caption{Patch antenna MRA}
         \label{fig:patch}
     \end{subfigure}
    \begin{subfigure}[b]{0.45\textwidth}
         \centering
         \includegraphics{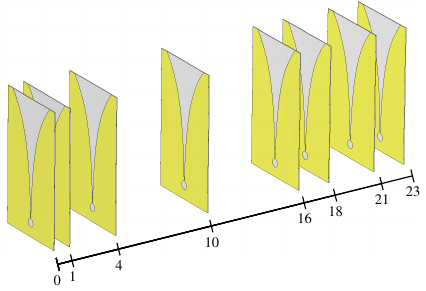}
        \vspace{-0.3cm}
         \caption{Vivaldi antenna MRA}
         \label{fig:vivaldi}
     \end{subfigure}
    \caption{Illustration of the studied $8$ element sparse arrays on Rogers RO4350B substrate. The element positions are multiples of $\lambda/2$ and the total aperture size is $11.5\lambda$ with $\lambda$ denoting the wavelength.}
    \vspace{-5mm}
    \label{fig:arrays}
\end{figure}

Based on above, there is a clear gap in the %
literature in terms of characterizing the DoA estimation performance of sparse arrays when considering 
realistic complex element gain patterns, directivity, and nonidealities including mutual coupling and variations due to manufacturing tolerances. 

In this paper, we seek to fill this important gap, especially focusing on challenging low-SNR scenarios with closely spaced sources where sparse arrays and directive antenna elements can drastically improve resolution. We study the effects of realistic antenna element types and their properties on DoA estimation capabilities and 
range of operable 
SNR levels. 
To this end, we consider uniform and sparse arrays with omnidirectional, patch and Vivaldi elements, and harness realistic EM simulations for their accurate modeling and characterization. Concrete examples of such sparse arrays are illustrated in Fig.~\ref{fig:arrays}. Importantly, mutual element coupling and individual element directivity properties are all taken into account when evaluating the performance of widely deployed DoA estimation algorithms, particularly MUSIC \cite{music}. 
The SNR improvements and sensing angle limitations with each antenna element type are 
assessed. The emphasis is on estimating DoAs of incoming signals in the low SNR regime. This is justified because the largest differences between antenna element types appear near the detection threshold, where improvements in SNR yield the highest gain in sensing sensitivity. Furthermore, the effect of nonidealities like mutual coupling and calibration errors due to manufacturing are considered to characterize the real-world performance in more detail.

\textls[-4]{ 
Our numerical results show that the utilization of directional antenna elements and sparse arrays allows substantially improved accuracy, compared to an  
omnidirectional  
ULA when estimating the directions of two closely spaced  
sources. Furthermore, the corresponding effective SNR gains are shown to translate into 4 to 15-fold increases in the maximum passive sensing distance between the sources and the sensor. Overall, the results show that with EM-simulated sparse arrays, patch and Vivaldi elements yield the best sensitivity for 120$^\circ$ and 80$^\circ$ field of views, respectively. The improvement is mainly due to directive elements and larger (sparse) array aperture with also nonidealities having a notable effect on the field-of-view and sensitivity.}

\vspace{-1mm}
\section{Background and Fundamentals}
\vspace{-0.5mm}
\subsection{Spatial signal model and notations}
First, the signal model for a linear antenna array is defined assuming $L$ uncorrelated far-field signals ${s_l}(t),l=1,\ldots,L$, and $N$ sensors or antenna elements. The received signal snapshot vector $\vec{x}(t)\in \mathbb{C}^{N}$ at instant $t$ is given by \cite{vantrees2002optimum}
\begin{equation}
    \vec{x}(t) = \vec{A}(\boldsymbol{\phi} ) \vec{s}(t) + \vec{n}(t) = \sum_{l=1}^L \vec{a}(\phi_l){s_l}(t) + \vec{n}(t),
    \label{eq:signal_model}
\end{equation}
where $\vec{s}(t)\in \mathbb{C}^{L}$ is the corresponding source signal vector, and $\vec{n}(t)\in \mathbb{C}^{N}$ is a noise vector. %
The matrix $\vec{A}(\boldsymbol{\phi} )\in \mathbb{C}^{N\times L}$ denotes the array manifold matrix, constructed from the $L$ steering vectors $\vec{a}(\phi_l)\in \mathbb{C}^{N}$ corresponding to source DoAs $\phi_l\in [-\tfrac{\pi}{2},\tfrac{\pi}{2}], l=1,\ldots,L$. The classical definition assumes ideal omnidirectional antennas. However, to also take into account the effect of the complex antenna element gain patterns, element-wise complex responses $g_n(\phi_l)$ are modeled, applicable for a given emitter polarization. Thus, steering vector $\vec{a}(\phi_l)$ may be written as
\begin{equation}
    \vec{a}(\phi_l)\!=\!\left[g_1(\phi_l)e^{-j\frac{2\pi p_1}{\lambda}\sin \phi_l} ,\cdots, g_N(\phi_l)e^{-j\frac{2\pi p_N}{\lambda}\sin \phi_l}\right]^T,
\end{equation}

where $\lambda$ denotes the wavelength and $p_n$ the physical location of the $n$th antenna element.
In the ideal case, $g_n(\cdot) = 1$ for all elements and directions.

\vspace{-0.5mm}
\subsection{Sparse arrays and high-resolution DoA estimation methods}

Sparse arrays provide multiple advantages over uniform arrays, such as enhanced identifiability and resolution \cite{amin2024sparse}. Well-known sparse array configurations include the \textit{nested array} (NA) \cite{NA} and \textit{minimum redundancy array} (MRA) \cite{MRA}. The MRA constitutes an optimal sparse array in the sense that it provides the narrowest main lobe (or largest aperture) for a fixed number of sensors, subject to its inter-sensor spacings constituting all integer multiples of $\lambda/2$ up to the aperture of the array. Finding MRAs becomes computationally challenging as the number of sensors $N$ grows. However, solutions are known for values of $N$ that are useful in practice \cite{MRA_perfect}. Hence, this work focuses on the MRA with $N=8$, which is benchmarked against the $8$-sensor ULA.

\textls[-4]{
Among the plethora of DoA estimation methods \cite{vantrees2002optimum}, arguably the simplest is the classical matched filter beamformer. However, it suffers from poor 
resolution and is not suitable when the emitters are closely spaced in the angular domain.\textit{ Maximum likelihood} (ML) methods \cite{ML}, in turn, provide high resolution capabilities with the expense of notable computational complexity.} 
\textls[-4]{
Subspace methods, including 
the widely deployed MUSIC algorithm \cite{music}, strike a balance between resolution and complexity, making them highly appealing in practice. 
MUSIC estimates the DoAs by decomposing the received signal $\vec{x}(t)$ into signal and noise subspaces, over observation instances $t=0,\ldots,T-1$. In classical element-space MUSIC, the DoA estimates are found by searching for peaks in the so-called MUSIC pseudospectrum, which at trial angle ${\phi_\text{tr}}$ is given by}
\begin{equation}
    P_{\mathrm{MU}}(\phi_\text{tr} ) = \frac{1}{\vec{a}^*(\phi_\text{tr} )\vec{V}_\mathrm{N} \vec{V}^*_\mathrm{N}\vec{a}(\phi_\text{tr})}.
    \label{eq:music}
\end{equation}
Here %
$(.)^*$ denotes conjugate transpose and $\vec{V}_\mathrm{N}\in\mathbb{C}^{N\times (N-L)}$ is the matrix of eigenvectors that span the noise subspace, see \cite{music} for details. 
Sparse arrays can further leverage the uncorrelated source signal assumption (given sufficiently many snapshots $T$) and employ a variant of MUSIC in a ``virtual'' element domain \cite{NA} to resolve more sources than sensors---a feat not possible for ULAs. For simplicity, however, this paper focuses on the canonical  
element-space MUSIC as the baseline DoA estimation approach.

\section{EM Modeling and Methodologies}

For the rest of this paper, we assume an array with $N=8$ elements which can be considered realistic and implementable 
with commercial multi-channel receivers. Furthermore, the MRA configurations are as shown in Fig. \ref{fig:arrays} for both studied antenna element types, i.e., patch and Vivaldi. Finally, we focus our study on 10 GHz in the middle of the X-band, as a concrete practical example -- however, we also note that the antennas could be scaled to virtually any frequency. 
\subsection{Radiation patterns and receiver SNR}

\begin{figure}
\centering
\begin{subfigure}[b]{0.24\textwidth}
         \centering
         \includegraphics{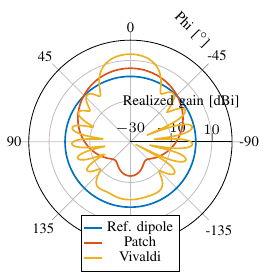}
         \vspace{-0.5cm}
         \caption{Magnitudes of element patterns}
         \label{fig:pattern_amplitude}
     \end{subfigure}
\begin{subfigure}[b]{0.24\textwidth}
         \centering
         \includegraphics{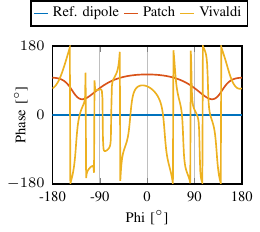}
         \vspace{-0.5cm}
         \caption{Phases of element patterns}
         \label{fig:pattern_phase}
     \end{subfigure}
    \caption{EM simulated antenna element patterns.}
    \vspace{-3mm}
    \label{fig:patterns}
\end{figure}

\textls[-12]{Complete EM simulations are carried out to accurately characterize 
the complex far-field radiation patterns, including phase and amplitude.
Antenna types selected for EM simulations include an ideal 
half-wave dipole, a typical PCB patch antenna, and Vivaldi antenna \cite{vivaldi}, \cite{vivaldi_parameter}. The patch and Vivaldi antennas are modeled with EM-simulations using CST Studio Suite on a realistic Rogers RO4350B PCB substrate at 10\,GHz. Fig. \ref{fig:arrays} illustrates the simulated antenna elements in an 8 element MRA configuration. Fig. \ref{fig:patterns} presents the obtained individual element amplitude and phase radiation patterns with the reference dipole having a constant omnidirectional gain of 2.15\,dBi, the patch antenna having a maximum gain of 8\,dBi while the Vivaldi antenna reaching a maximum gain of 13\,dBi. The patch element has a quite flat phase response, but Vivaldi has a highly varying phase response, indicating a higher sensitivity to phase calibration errors.}

Furthermore, to evaluate the SNR for each sparse array configuration and antenna element type in a fair manner, element-level isotropic SNR is defined as
\begin{equation}\label{Eq:SNR}
    \mathrm{SNR} \triangleq \frac{\lambda^2}{4\pi \eta_0}\frac{|\vec{E}|^2}{P_{\mathrm{N}}},
\end{equation}
\textls[-1]{which assumes a 0 dBi gain receiving antenna and is defined by the source electric field strength $|\vec{E}|$ at the receiver location independent from the array and geometry type, receiver noise power $P_{\mathrm{N}}$, %
and free-space wave impedance $\eta_0$.
In this expression, the noise level stays constant in all the cases compared. The receiver noise is assumed to be dominant, and the antenna noise temperature is assumed to be equal in all cases.}

\subsection{Mapping SNR values to source sensing distance} \label{Sec:SNRdistance}

To
highlight the differences among the geometry and antenna element options along the numerical results in the upcoming Section IV, the different SNR values are mapped to different free-space path loss distances corresponding to a fixed power source while maintaining a fixed average DoA estimation error. An emitter with the power of $P_t$ and gain $G_t$ towards the sensor is assumed to radiate a source electric field density of $|\vec{E}|=\sqrt{\frac{P_t G_t}{4\pi r^2}}$ at the receiver location. By combining the electric field density with Eq. \eqref{Eq:SNR}, the one-way (passive) sensing distance of the source can be solved and expressed as

\begin{equation}\label{Eq:r}
    r=\sqrt{\frac{P_t G_t \lambda^2}{16 \pi^2 \eta_0 P_{\mathrm{N}} \times \mathrm{SNR}}}.
\end{equation}

Using the ULA with ideal reference elements as a baseline case, the relative improvement in the coverage distance can then be expressed as $r=r_0\frac{\mathrm{SNR_0}}{\mathrm{SNR}}$, where $\mathrm{SNR_0}$ and $\mathrm{SNR}$ are the reference ULA and the studied array SNR levels needed for achieving the same angular RMSE in DoA estimation.

\subsection{Modeling physical nonidealities}\label{Sec:modeling_nonidealities}

\textls[-4]{In actual physical arrays, mutual coupling and other platform-based interactions affect the electromagnetic behavior and thereby the sensing performance. Thus, in addition to single-element EM simulations, this work also conducts \emph{full-array EM simulations} 
that include the whole physical antenna array structure in a single EM simulation model. Hence, these complete EM simulation results include all interactions between the antennas, and are reported separately in the upcoming Section~\ref{sec:results}.}

Additionally, in the real world, the array manifold 
$\vec{a}(\cdot)$ is not known perfectly, which impacts DoA estimation performance. If each antenna array is calibrated separately, in theory the measured element gain and phase patterns include all imperfections. However, practical arrays often include errors, especially in the phase calibration measurement, which can be modeled as additive noise in the phase pattern. In mass production, typically not all individual arrays are measured, leading to physical variations in dimensions and materials causing errors. When an RF-substrate is assumed, most variation is likely to be in the PCB stack-up with the promise of PCB thickness tolerance from some manufacturers being only $\pm10\%$. Thus, in this paper, such PCB thickness variation is also included and EM simulated as a dimensional error.

\section{DoA Estimation Results and Analysis}
\label{sec:results}

\subsection{Assumptions}
\textls[-1]{Target emitters are modeled as noise sources, using Gaussian distributed complex random signals that well approximate modern communication modulations. 
Thermal noise is also modeled as circular complex Gaussian.

We assume $T=50$ temporal samples (snapshots) 
for a realistic assessment and comparison in the case of LPD type of waveforms.
When estimating the DoAs from the MUSIC pseudospectrum in \eqref{eq:music}, the number of emitters is assumed to be known. This assumption is made to focus the study to the DoA estimation capability only. 
The root mean square error (RMSE) is used as a main error criterion in comparisons.
Finally, in the baseline simulations in the upcoming \cref{sec:individualEM}, it is assumed that the array manifold $\vec{a}(\cdot)$ is perfectly known. 
Hence, there are no errors in the assumed antenna geometry or in the array calibration. Calibration errors and other nonidealities are studied in \cref{sec:fullEM}, along the full-array EM simulations.}

\subsection{Results with invidual element EM simulations}\label{sec:individualEM}

\begin{figure*}
    \centering
    \begin{subfigure}[b]{0.49\textwidth}
         \centering
         \includegraphics{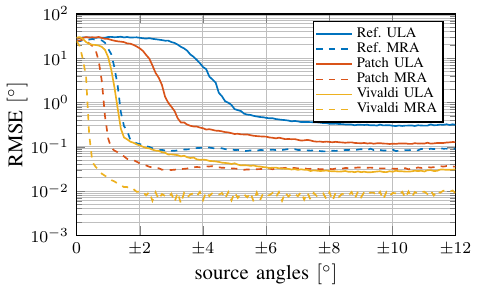}
         \vspace{-2mm}
         \caption{With individually EM-simulated antenna elements}
        \label{fig:average_error_5db}
    \end{subfigure}
    \begin{subfigure}[b]{0.49\textwidth}
         \centering
         \includegraphics{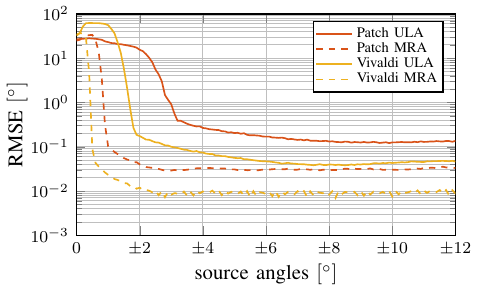}
         \vspace{-2mm}
         \caption{With full-array EM-simulation}
        \label{fig:average_error_5db_cosim}
    \end{subfigure}
    \caption{DoA estimation RMSEs with $L=2$ symmetric sources at small angular separations and element level isotropic SNR of --5\,dB. %
    The sparse array geometry and directional elements 
    provide improved performance.
    }
    \vspace{-2mm}
\end{figure*}

\begin{figure*}
    \centering
     \begin{subfigure}[b]{0.49\textwidth}
         \centering
         \includegraphics{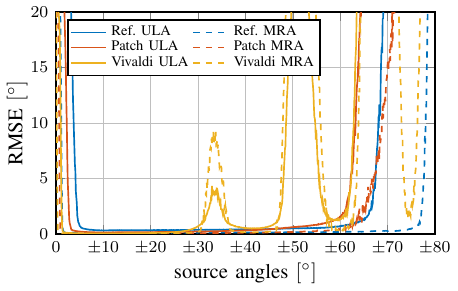}
         \vspace{-2mm}
         \caption{With individually EM-simulated antenna elements}
    \label{fig:average_error_5db_large}
    \end{subfigure}
    \begin{subfigure}[b]{0.49\textwidth}
         \centering
        \includegraphics{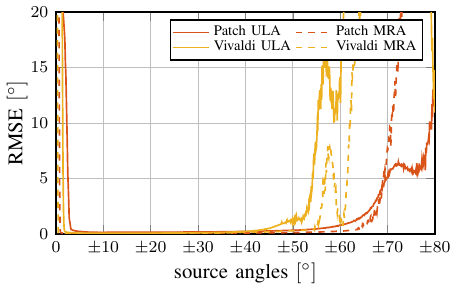}
         \vspace{-2mm}
         \caption{With full array EM-simulation}
    \label{fig:average_error_5db_large_cosim}
    \end{subfigure}
    \caption{DoA estimation RMSEs with $L=2$ symmetric sources for broader range of source angles at element level isotropic SNR of --5\,dB.}
    \vspace{-3mm}
\end{figure*}

First, DoA estimation performance is assessed with calibrated antenna arrays consisting of elements with all having an identical isolated EM-simulated radiation pattern. This is to first characterize the performance without the nonideal effects of non-identical radiation patterns, mutual coupling and other interactions between elements of the array. Fig.~\ref{fig:average_error_5db} shows the DoA estimation RMSEs when varying the source angles of two symmetric sources around zero degrees at the broadside of the array with the MUSIC algorithm at element level isotropic SNR of --5\,dB. It can be seen that 
resolving closely spaced sources is challenging for the ULA, 
while the MRA significantly enhances resolution even when using omnidirectional (reference) antenna elements. 
Directional elements further improve the resolution (both in case of the MRA and ULA) by reducing the minimum angular separation needed to reliably resolve the source directions. Assuming an example target RMSE of 0.4$^\circ$ or less, the improvement is from $13.2^\circ$ down to $2.8^\circ$ with directional Vivaldi antenna elements in ULA compared to reference ULA. 
The best performance is achieved when using both MRA sparse array and Vivaldi antennas, with minimum angle separation reduced to $0.8^\circ$. 

\begin{table}[t]
    \centering
    \caption{Comparison of DoA estimation RMSEs with $L=2$ symmetric sources at $\pm10^\circ$ and $-5\, \mathrm{dB}$ element-level isotropic SNR (wrt. to ref. ULA)}
    \label{tab:10degree_results}
    \setlength{\tabcolsep}{0.6em}
    \begin{tabular}{|>{\centering\arraybackslash}p{2cm}|>{\centering\arraybackslash}p{0.6cm}>{\centering\arraybackslash}p{0.8cm}|>{\centering\arraybackslash}p{0.6cm}>{\centering\arraybackslash}p{0.8cm}|>{\centering\arraybackslash}p{0.7cm}>{\centering\arraybackslash}p{0.8cm}|}
        \hline

        Type & \multicolumn{2}{|c|}{Reference}& \multicolumn{2}{|c|}{Patch} & \multicolumn{2}{|c|}{Vivaldi}\\
        \hline
        ULA& $0.30 ^\circ$ & (0\%) & $0.12 ^\circ$ & (-61\%)\textbf{}& $0.027 ^\circ$ & (-91\%) \\
         MRA& $0.053 ^\circ$ & (-71\%) & $0.032 ^\circ$ & (-89\%) & $0.0089 ^\circ$ & (-97\%) \\
        \hline
        ULA w/ full-EM& & & $0.13 ^\circ$ & (-58\%)& $0.043 ^\circ$ & (-86\%) \\
         MRA w/ full-EM& & & $0.030 ^\circ$ & (-90\%) & $0.0076^\circ$ & (-97\%) \\
        \hline
    \end{tabular}
    \vspace{-5mm}
\end{table}
Table \ref{tab:10degree_results} compares the DoA estimation RMSEs of the different array and antenna combinations in \cref{fig:average_error_5db}, along with the relative improvements compared to the uniform array with the reference dipoles. Here, two symmetric sources at $\pm10^\circ$  ($20^\circ$ separation) are assumed. As seen through the additional results in \cref{fig:average_error_5db_large} which shows the DoA estimation performance for wider range of source angles, the error levels remain essentially constant up to source angles of around $\pm25^\circ$ ($50^\circ$ separation), after which the nulls in the Vivaldi element radiation pattern increase the errors locally around the null angles (cf. also Fig. \ref{fig:patterns}). 
The field of view is widest in case of the MRA with ideal dipole antenna elements. With patch elements, the field of view decreases compared to dipoles, both for the ULA and MRA, since the element pattern limits the useful angular range.

\begin{figure*}
    \centering
    \begin{subfigure}[b]{0.49\textwidth}
         \centering
          \includegraphics{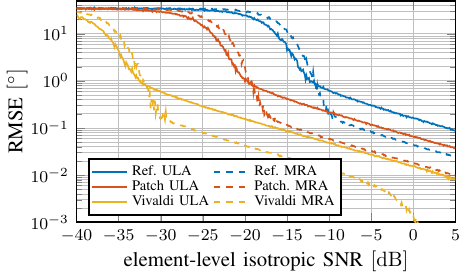}
         \vspace{-2mm}
         \caption{With individually EM-simulated antenna elements}
         \label{fig:average_error_SNR_sweep}
    \end{subfigure}
    \begin{subfigure}[b]{0.49\textwidth}
         \centering
          \includegraphics{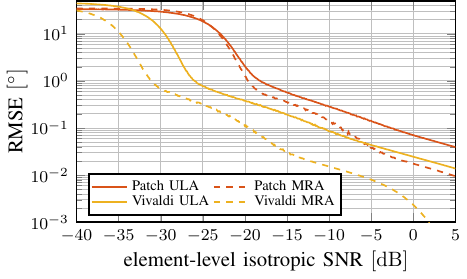}
         \vspace{-2mm}
         \caption{With full array EM-simulation}
         \label{fig:average_error_SNR_cosim}
    \end{subfigure}
    \caption{RMSE in DoA estimates with varying element-level isotropic SNR with $L=2$ symmetric sources at $\pm10^\circ$ source angles highlighting the improvement from sparse array and directional elements.}
    \vspace{-4mm}
\end{figure*}

\textls[-4]{Next, Fig. \ref{fig:average_error_SNR_sweep} shows the RMSE performance as a function of the element-level isotropic SNR for two sources at $\pm 10^\circ$ $(20^\circ$ separation) with respect to the broadside of the array. 
With RMSE levels lower than around 0.7--0.8$^\circ$, both the sparse array and the directional antenna give a significant advantage in the SNR required to achieve a given error level. At SNR levels below $-10$\,dB, the reference sparse array is slightly worse than the reference ULA. However, directional antenna elements still yield a significant reduction in RMSE at much lower SNR values.}

\subsection{Results with full-array EM simulations
}\label{sec:fullEM}
\vspace{-0.5mm}

\textls[-2]{Next, corresponding results with complete full-array EM simulations are provided where all antenna elements are included in one EM-simulation model to take into account nonidealities, such as non-identical radiation patterns, cross-coupling between the antennas and other effects from the platform. These can thus be considered the main and most realistic results.}

\textls[-2]{Fig. \ref{fig:average_error_5db_cosim} shows the RMSE performance for small source angles and corresponding angular separations. We again observe that resolving the closely spaced sources is challenging for the baseline ULA approach. Furthermore, both the sparse MRA array and the Vivaldi antenna elements significantly increase the resolution, reducing the minimum anglular separation needed to reliably resolve the source directions. Assuming an example target RMSE of $0.4^\circ$, the improvement is from $6.4^\circ$ down to $1.8^\circ$ with patch elements and from $3.6^\circ$ down to $1^\circ$ with directional Vivaldi elements when changing from ULA to MRA. Table \ref{tab:10degree_results} collects the results with $\pm10^\circ$ source angles with full EM-simulations, allowing also to compare against the previous results with individual element EM modeling. The results are similar to the results with individual antenna element patterns, with the mutual coupling slightly reducing the benefit from directional elements and sparse array geometry in selected cases -- however, the degradation is essentially negligible at 
the considered 
SNR of $-5$\,dB and
source separation of $20^\circ$.} 

\textls[-2]{Fig. \ref{fig:average_error_5db_large_cosim} shows then the corresponding RMSE results for wider range of source angles. We observe that with mutual coupling, the effect of the first Vivaldi antenna pattern null is significantly reduced, increasing the real usable field of view up to $\pm40^\circ$ %
with the ULA, and more, up to $\pm50^\circ$, with the sparse array. Fig. \ref{fig:average_error_SNR_cosim} shows the RMSE with respect to SNR in the full array EM-simulation. In case of the ULA, the performance with patch elements remains closely the same as in the individual element simulation, whereas with Vivaldi elements, performance degrades the most, especially at SNR values below $-$15 dB. 
The performance of the patch and Vivaldi MRA 
closely corresponds to  the individual EM-simulations. This is consistent with sparse arrays being less susceptible to mutual coupling \cite{coupling}.
}

\begin{figure*}
    \centering
     \begin{subfigure}[]{0.49\textwidth}
         \centering
         \includegraphics{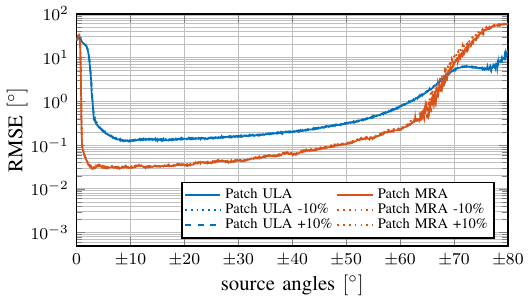}
         \vspace{-6mm}
         \caption{Patch elements}
         \label{fig:y equals x}
     \end{subfigure}
     \begin{subfigure}[]{0.49\textwidth}
         \centering
         \includegraphics{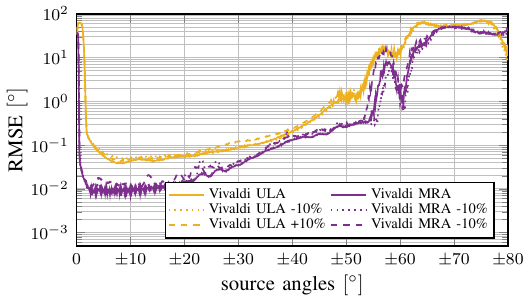}
         \vspace{-6mm}
         \caption{Vivaldi elements}
         \label{fig:three sin x}
     \end{subfigure}
    \caption{DoA estimation RMSEs building on complete full-array EM-simulations including small physical perturbations caused by percentual PCB thickness errors. Two symmetric sources at element level isotropic SNR of --5\,dB.}
    \vspace{-6mm}
    \label{fig:dimension_error}
\end{figure*}

\textls[-1]{Finally, the effects of manufacturing tolerances are studied as described in \cref{Sec:modeling_nonidealities}. Fig. \ref{fig:dimension_error} shows the effect of PCB thickness variation on the DoA estimation peformance. 
This essentially characterizes the sensitivity of the antenna arrays to model mismatch, as the true array manifold is not anymore perfectly known.
The results with 10\%
PCB thickness errors indicate that the patch antenna is very robust  to such dimensional errors as the results for all patch arrays are almost the same. With Vivaldi arrays, a very slight increase in RMSE can be seen with dimensional errors, but again the effects are minor.}

\vspace{-2.0mm}
\subsection{Differences in free-space path loss distances}
\vspace{-0.5mm}
\textls[-2]{The fundamental SNR performances of the different array configurations and element types are as presented in Figs. \ref{fig:average_error_SNR_sweep} and \ref{fig:average_error_SNR_cosim}. Now, by assuming a fixed RMSE requirement, the differences in coverage distances can be calculated for different array configurations and element types according to Section~\ref{Sec:SNRdistance}. 
By using the reference ULA coverage distance $r_0$ as a reference, the improvements in coverage distance are calculated and shown in Table \ref{tab:detection} for a challenging RMSE target of
$0.1^\circ$. The improvements are valid for an ideal line-of-sight channel. In practice, radio horizon limits the maximum distance.}

It is seen that without mutual coupling, using a MRA leads to 3 or 4 times longer coverage distance compared to a ULA with each antenna type. Furthermore, when changing to directive patch elements from the reference elements, 2- to 3-fold increase in distance is observed for both ULA and MRA, while 10-fold increase is obtained with even more directive Vivaldi elements. Notably, with a sparse array and Vivaldi elements, over 30-fold increase in coverage distance is seen compared to a ULA with the reference dipole elements.

Furthermore, with mutual coupling and non-identical patterns included, the patch array performance has almost no change in ULA configuration but the MRA is more sensitive with coverage distance decreasing from $7.9 r_0$ to $4.5 r_0$. For Vivaldi elements, both ULA and MRA are sensititive, with the distance with Vivaldi MRA decreasing from $35.5 r_0$ to $15 r_0$. The exact degradations depend on the SNR operating region in Figs. \ref{fig:average_error_SNR_sweep} and \ref{fig:average_error_SNR_cosim}, but generally especially with MRA, the non-identical patterns and mutual coupling need to be taken into account when evaluating the arrays in the low SNR regime.

\begin{table}[t]
    \centering
    \caption{SNR differences mapped to equivalent one-way free-space path loss distances with $0.1^\circ$ maximum RMSE for two symmetric sources at angles of $\pm10^\circ$}
    \label{tab:detection}
    \centering
    \begin{tabular}{|c|ccc|}
        \hline
        Type & Reference& Patch & Vivaldi\\
        \hline
        ULA& $r_0$ & $2.5 r_0$ & $11.2 r_0$\\
         MRA& $3.5 r_0$ & $7.9 r_0$ & $35.5 r_0$  \\
        \hline
        ULA w/ full-EM&  & $2.4 r_0$  & $5.0 r_0$  \\
         MRA w/ full-EM&  & $4.5 r_0$  & $15.0 r_0$  \\
        \hline
    \end{tabular}
    \vspace{-5mm}
\end{table}
\vspace{-.2cm}
\subsection{Closing example with more sources than sensors, $L>N$}

\begin{figure}
    \centering
    \includegraphics{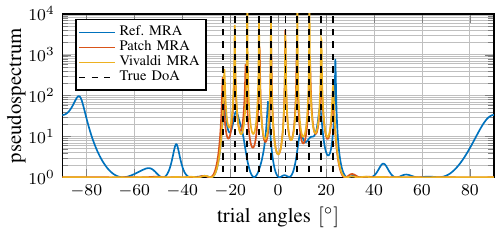}
    \vspace{-2mm}
    \caption{Coarray MUSIC pseudospectrum with $L=10$ sources ($>N=8$), 
    $\mathrm{SNR}=-25\;\mathrm{dB}$ and $T=1024$ snapshots.}
    \vspace{-6mm}
    \label{fig:presudospectrum 10_sources}
\end{figure}

Finally, a case with larger number of sources, $L$, is showcased to highlight the 
ability of sparse arrays (with directive elements) to identify more sources than sensors. 
The number of sources is set to $L=10$. As this is larger than the number of sensors $N=8$, coarray MUSIC \cite{coarray_music} is employed instead of element-space MUSIC. 
Fig \ref{fig:presudospectrum 10_sources} shows a realization of the pseudospectrum in case of the MRA with individual element EM simulations. 
Both patch and Vivaldi antenna elements can correctly identify all sources. This is a clear improvement compared to the array with reference dipole, as it has false peaks and is not able to determine 4 out of 10 sources due to its lower antenna gain in the source directions. 
In addition, while not observable directly 
in the 
pseudospectrum, 
sources close to boresight have lower error when using Vivaldi elements, whereas sources further from boresight have lower error using patch elements.
This is 
consistent with 
Fig \ref{fig:average_error_5db_large}. Results with the ULA are not shown, as an $N$-sensor ULA cannot 
identify more than $N-1$ sources.

\vspace{-3mm}
\section{Conclusion}
This paper investigated high-resolution DoA estimation of sparse arrays -- particularly the MRA -- while benchmarking against ULA when employing practical directive 
antenna element radiation patterns.
The impacts of complex antenna element gain patterns and nonidealities, including mutual coupling and manufacturing tolerances, were modeled and studied via realistic EM simulations, considering a practical example case of an 8-element array at 10\,GHz.
Both the directive antennas and the sparse array were shown to provide significant performance improvements, also when taking account of the physical nonidealities. 
In the important study case of estimating two closely-spaced sources, deploying sparse MRA was shown to reduced the DoA estimation RMSE by up to 90\% and 97\%, for patch and Vivaldi elements, respectively, when compared to a reference ULA with dipole elements.
Furthermore, for a fixed DoA estimation RMSE requirement of 0.1$^\circ$, a 4.5-times improvement in detection distance was obtained with a patch MRA, and even a 15-fold increase with a Vivaldi MRA, compared to reference ULA. However, with Vivaldi antenna elements, nulls in the radiation pattern limit the usable field of view to around $100^\circ$ sector while the patch elements allow sectors up to $120^\circ$ at $-$5 dB element-level isotropic SNR, albeit with a lower gain in said sector. 

In conclusion, patch array elements strike a good balance between field of view and antenna gain. 
Sparse array geometries provide further resolution gains at low SNR. More directive antennas, like Vivaldi, have even higher potential for SNR-gain if limiting the field of view to $100^\circ$ sector is acceptable.

\bibliographystyle{IEEEtran}
\bibliography{RadarConf2025}

\end{document}